\documentclass[a4paper,11pt]{article}
\usepackage{jheppub} 
\usepackage{lineno}

\title{Behaviour of $n$-point scattering amplitudes at high energies}

\author{Shreya Shrivastava}
\affiliation{Department of Physics, Indian Institute of Technology Bombay, India}

\emailAdd{shreyakshrivastava@gmail.com}

\abstract{ We study the on-shell scattering amplitudes in quantum gravity for high-energy collisions in the eikonal approximation. We first evaluate the $n$-loop 2-particle scattering amplitude in the high energy and low momentum transfer limit. We do so in a symmetrized manner by finding the contributions of each of the particle worldines to the scattering amplitude and gluing them together via the $n$ intermediate particle exchanges. In this limit on applying the eikonal approximation and summing over all $n$-loop Feynman diagrams we obtain a closed form for the 2 particle scattering amplitude. Finally, we extend this approach to obtain a generalized eikonal approximation for $N$-particle scattering at high energies and small momentum transfers. The generalised form of the scattering amplitude can then be used to evaluate the bound states of the system.}

\begin{document}
\maketitle
\flushbottom
\section{Introduction}
In light of 't Hooft's work which draws a relation between Veneziano amplitudes in string theory and high energy two-particle scattering in quantum gravity and string theory ~\cite{tHooft:1987vrq}, several calculations have been done to establish a similar connection between the two pictures. One such approach has been done by Muzinich and Soldate ~\cite{PhysRevD.37.359} where four-point string amplitudes at large center-of-mass energy $\sqrt{s}$
and fixed momentum transfer $q =-t$ is related to 2-particle scattering amplitude obtained by summing multiple Reggeized graviton exchange in the eikonal approximation in D space-time
dimensions. Additionally, the expression obtained for 2-2 scattering in QED in the eikonal approximation has been used to determine the relativistic Balmer formula in ~\cite{PhysRevD.1.2349}. However the aforementioned works have been limited to 2-particle scattering amplitudes, due to the more involved calculations required for N-particle scattering. In this paper we develop a symmetric approach to relativistic eikonal calculations resulting in a straightforward generalization to higher point amplitudes. We further confirm the result obtained by H. and E. Verlinde ~\cite{Verlinde_1992} stating that N-particle amplitudes describing the gravitational
scattering of the matter particles factorize into the product of
two-particle amplitudes.

\section{Eikonal scattering amplitudes}
We will now study the the behaviour of scattering amplitudes for $N$ scalar particles of mass $m$, at high energies and small momentum transfers. We specifically look at the case of scattering amplitudes in linearised gravity.
\vspace{2mm}\\
The signature of the Minkowski metric being considered henceforth is $\eta_{\mu\nu}=\text{diag}(-1,1,1,1)$. The action for Einstein gravity coupled to a real scalar field expanded about a Minkowski background, $g_{\mu \nu}=\eta_{\mu \nu}+h_{\mu \nu}$ upto leading order in $h_{\mu \nu}$ is given by ~\cite{Kabat_1992}
\begin{align*}
S= & \int \mathrm{d}^4 x \frac{1}{16 \pi G} \frac{1}{8} h_{\alpha \beta}\left[\eta^{\alpha \gamma} \eta^{\beta \delta}+\eta^{\alpha \delta} \eta^{\beta \gamma}-\eta^{\alpha \beta} \eta^{\gamma \delta}\right] \square h_{\gamma \delta}+\frac{1}{2} \phi\left(\square-m^2\right) \phi \\
& +\frac{1}{2} h_{\mu \nu}\left[\partial^\mu \phi \partial^\nu \phi-\frac{1}{2} \eta^{\mu \nu}\left(\partial_\lambda \phi \partial^\lambda \phi+m^2 \phi^2\right)\right] .
\end{align*}
The corresponding Feynman rules are 
\begin{itemize}
    \item Scalar propagator: $i\Delta=-\frac{i}{p^2+m^2-i\epsilon}$
    \begin{figure}[h]
    \centering
    \includegraphics[scale=0.75]{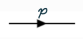}
    \label{fig:my_label}
\end{figure}
    \item Graviton propagator: $iD^{\alpha\sigma\beta\tau}=-i\frac{16\pi G}{k^2-i\epsilon}(\eta^{\alpha\sigma}\eta^{\beta\tau}+\eta^{\alpha\tau}\eta^{\beta\sigma}-\eta^{\alpha\beta}\eta^{\sigma\tau})$
    \begin{figure}[h]
    \centering
    \includegraphics[scale=0.75]{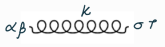}
    \label{fig:my_label}
\end{figure}
    \item Interaction vertex:$\frac{i}{2}(p_\alpha p'_\beta+p_\beta p'_\alpha-\eta_{\alpha\beta}(p\cdot p'+m^2))$
    \begin{figure}[h]
    \centering
    \includegraphics[scale=0.75]{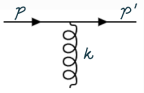}
    \label{fig:my_label}
\end{figure}
\end{itemize}
In evaluating the vertex factors, we ignore the recoil of the matter field,
\[p_{\mu}p'_{\nu}+p_{\nu}p'_{\mu}-\eta_{\mu\nu}(p\cdot p'+m^2)\approx2p_{\mu}p_{\nu}\]
In the matter propagators, we ignore $k^2$ relative to $p\cdot k$,
\[\frac{1}{(p+k)^2+m^2-i\epsilon}\approx \frac{1}{2p\cdot k-i\epsilon}\]

\subsection{Four point scattering amplitude}
The $n$ loop diagram corresponding to four point scattering will look as follows,
\begin{figure}[H]
    \centering
    \includegraphics[scale=0.9]{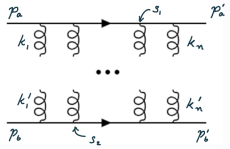}
    \label{fig:my_label}
    \caption{Visualisation of any n-loop diagram representing all possible ladder and cross-ladder diagrams obtained by all combinations of connections made between graviton emission and absorption points}
\end{figure}
The contribution of this diagram to the four point scattering amplitude will be
\begin{equation}
    -i\mathcal{M}_{n}=(-ig)^{2n}\int \prod_{i=1}^n\frac{d^4k_j}{(2\pi)^4}\Tilde{\Delta}(k_j)\times I\times (2\pi)^4\delta(q-\sum_{i=1}^n k_i)
\end{equation}
where $\Tilde{\Delta}(k)=i/(k^2-\mu^2-i\epsilon)$ is the meson propagator and $q=p_a-p'_a=-(p_b-p'_b)$ and $I$ is a sum of products of propagators associated with the propagation of particles $a$ and $b$. Also, $g^2=16\pi G\gamma(s)$, where $\gamma(s)=2(p_1\cdot p_2)^2-m^4$. Now, using the following inverse Fourier transformation,
\begin{equation}
\label{h}
    \Tilde{\Delta}(k_s)=-\frac{1}{k_s^2-i\epsilon}=\int \Delta(x) e^{-ik_s\cdot x}d^4 x
\end{equation}
for the $s^{\text{th}}$ term, we can rewrite $i\mathcal{M}_n$ as 
\begin{equation}
    i\mathcal{M}_n=-g^2\int d^4 x e^{-iq\cdot x}\Delta(x)(2\pi)^4\underbrace{\left(\int \prod_{i=1}^{n-1}\frac{d^4k_j}{(2\pi)^4}(ig)^2\Tilde{\Delta}(k_j)\times I \right)}_{(A)}
\end{equation}
Inorder to evaluate $I$ we first look at the contribution of the propagators from the worldline of $a$. Lets use the $\delta$  function in the above expression to eliminate $k_{s_1}$. For the simplest configuration with $k_i$ are emitted sequentially according to their index
\begin{equation}
\label{c}
    I_{s_1}^{a}=\Delta^a(p_a-k_1)\dots\Delta^a(p_a-k_1-\dots -k_{s_1-1})\\
    \times\Delta^a(p'_a+k_{n})\dots\Delta^a(p'_a+k_n+\dots +k_{s_1+1})
\end{equation}
where $\Delta^a(p)=i/(p^2-m_a^2-i\epsilon)$. 

\subsubsection{Eikonal approximation}
\label{sec:worldlines}
We now apply the following propagator approximation 
\begin{equation}
\label{a}
    (p\pm K)^2+m^2=\pm 2p\cdot K+K^2\approx\pm2p\cdot K
\end{equation}
on the mass shell, where we take the external momentum $p$ to be much larger than the partial sum of internal momenta $K$. However we lose certain symmetry properties of the propagator due to this approximation. By momentum conservation we have
\begin{equation}
\label{b}
    (p_a-k_1\dots -k_{s_1})^2=(p_a'+k_{s_1+1}\dots +k_n)^2
\end{equation}
\[\implies \Delta(p_a-k_1\dots -k_{s_1})=\Delta(p_a'+k_{s_1+1}\dots +k_n)\]
but on applying the aforementioned approximation \eqref{a} the LHS and RHS in \eqref{b} become $-2p_a\cdot (k_1+\dots+k_{s_1})$ and $2p_a\cdot(k_{s_1+1}+\dots+k_n)$ which are not equal anymore. To account for this symmetry we add a factor of $1/n$
\begin{equation}
\label{g}
    I\rightarrow \frac{1}{n}I
\end{equation}
Coming back to evaluating the contribution from the worldline of $a$ in the eikonal approximation, \eqref{c} simplifies to
\begin{equation}
    I^{(a) \text{eik}}_{s_1}=i^{n}[(a_1)^{-1}(a_1+a_2)^{-1}\dots (a_1+\dots+a_{s_1-1})^{-1}]\\
    \times [(a'_n)^{-1}(a'_n+a'_{n-1})^{-1}\dots(a'_{n}+\dots a'_{s_1+1})^{-1}]
\end{equation}
where $a_i=-2p_a\cdot k_i+i\epsilon$ and $a'_i=2p'_a\cdot k_i+i\epsilon$. However as $k_i$ are merely labels, the set of permutations $\pi_1$ of $(k_1\dots k_{s_1-1})$ and $\pi_2$ of $(k_{s_1+1}\dots k_n)$ leave $I^{(a) \text{eik}}_{s_1}$ unchanged. Therefore
\begin{equation}
\begin{aligned}
    \label{d}
    I^{(a) \text{eik}}_{s_1;sym}&=\frac{i^{n}}{(s_1-1)!(n-s_1)!}\sum_{\pi_1,\pi_2}[(a_1)^{-1}(a_1+a_2)^{-1}\dots (a_1+\dots+a_{s_1-1})^{-1}]\\
    &\times [(a'_n)^{-1}(a'_n+a'_{n-1})^{-1}\dots(a'_{n}+\dots a'_{s_1+1})^{-1}]\\
    &=\frac{i^{n}}{(s_1-1)!(n-s_1)!}[a_1a_2\dots a_{s_1-1}]^{-1}[a'_{s_1+1}\dots a'_n]^{-1}
\end{aligned}
\end{equation}
where in the last step we have used the identity proved in \ref{identity} to obtain the final expression.

\subsubsection{Gluing the worldlines}
From the above result we have the following contributions from the worldlines of $a$ and $b$. Let the internal momenta for the worldline of $b$ be denoted by $k'_i\in\{k_1\dots k_n\}$. Then
\begin{equation}
\begin{aligned}
    \label{e}
    I^{(a) \text{eik}}_{s_1;sym}
    &=\frac{i^{n}}{(s_1-1)!(n-s_1)!}[a_1a_2\dots a_{s_1-1}]^{-1}[a'_{s_1+1}\dots a'_n]^{-1}\\
     I^{(b) \text{eik}}_{s_2;sym}
    &=\frac{i^{n}}{(s_2-1)!(n-s_2)!}[b_1b_2\dots b_{s_2-1}]^{-1}[b'_{s_2+1}\dots b'_n]^{-1}
\end{aligned}
\end{equation}
where $b_i=2p_b\cdot k'_i+i\epsilon$ and $b'_i=-2p'_b\cdot k'_i+i\epsilon$ and we have chosen to rewrite $k'_{s_2}$ by momentum conservation. Now inorder to glue the two together, we must sum over all possible ways to interconnect each of the $(n-1)$ vertices (from where the internal momenta arise) between the two branches. Let $\Pi$ be the operation that glues two worldlines together, and can be visualised as
\begin{figure}[H]
    \centering
    \includegraphics[scale=1]{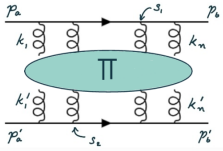}
    \label{fig:my_label}
    \caption{$\Pi$ gives the suitable combinatorial factor for number of ways to make the connections}
\end{figure}
\begin{equation}
\label{f}
    I(s_1,s_2)=I^{(a) \text{eik}}_{s_1;sym}(\Pi)I^{(b) \text{eik}}_{s_2;sym}
\end{equation}
From \eqref{e} we see that there are $s_1-1$ terms that are of the form of $a_i$. Suppose we select $r$ of these terms and pair them with $b_i$s, there are $\binom{s_1-1}{r}$ ways. Now we must pair the remaining $(s_2-1)$ $b_i$s with $a'_i$s, which can be done in $\binom{n-s_1}{s_2-1-r}$ ways. The left over terms can directly be paired among themselves i.e. $(s_1-1-r)$ of $(a_ib_i)$s and $(n-s_1-s_2+r+1)$ of $(a'_ib'_i)$s. Finally note that $k'_i$s was formed from the splitting of an arbitrary arrangement of the set $\{k_1,\dots,k_{s_1-1},k_{s_1+1},\dots,k_n\}$ into two complementary subsets. The degeneracy \footnote{As the internal momenta $k$s are labelled by dummy indices considering all possible arrangements of $k_i$s leads to the over counting.} eliminated by the $(s_2-1)!(n-s_2)!$ factor in \eqref{d} is now broken. This can be understood by the following case\\
\begin{figure}[h]
    \centering
    \includegraphics[scale=0.9]{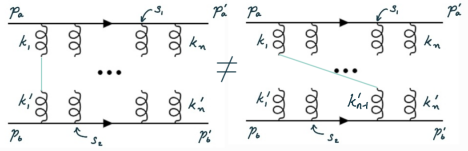}
    \label{fig:my_label}
    \caption{The left diagram with $k_1'=k_1$ and the right diagram with $k_1=k_{n-1}'$ are not equivalent diagrams}
\end{figure}\\
Hence we restore the factor $(s_2-1)!(n-s_2)!$. This gives
\begin{equation}
\begin{aligned}
\label{k}
    I=\sum_{s_1,s_2}I(s_1,s_2)
    &=\sum_{r,s_1,s_2}I^{(a) \text{eik}}_{s_1;sym}\frac{(s_1-1)!(n-s_1)!(s_2-1)!(n-s_2)!}{r!(s_1-r-1)!(s_2-r-1)!(n-s_1-s_2+r+1)!}I^{(b) \text{eik}}_{s_2;sym}\\
\end{aligned}
\end{equation}
Substituting the above expression for $I$ in \eqref{f}, $A$ factorises as follows,
\begin{equation}
\begin{aligned}
\label{i}
   A&=\sum_{r,s_1,s_2}\int\prod_{i=1}^{n-1} \frac{d^4k_i}{(2\pi)^4}(-ig)^2\Tilde{\Delta}(k_i)e^{ik_i\cdot x} I\\
    &=\sum_{r,s_1,s_2}\frac{1}{\alpha!\beta!\gamma!\delta!}u_1^\alpha u_2^\beta u_3^\gamma u_4^\delta\\
    &=\frac{(u_1+u_2+u_3+u_4)^{n-1}}{(n-1)!}=\frac{(i\chi)^{n-1}}{(n-1)!}
\end{aligned}
\end{equation}
where $\alpha=r,\beta=s_1-r-1,\gamma=s_2-r-1,\delta=n-s_1-s_2+r+1$ and 
\begin{align*}
    u_1&=g^2\int \frac{d^4k}{(2\pi)^4}\Tilde{\Delta}(k)e^{ik\cdot x}[ab]^{-1} &  u_2&=g^2\int \frac{d^4k}{(2\pi)^4}\Tilde{\Delta}(k)e^{ik\cdot x}[ab']^{-1}\\
    u_3&=g^2\int \frac{d^4k}{(2\pi)^4}\Tilde{\Delta}(k)e^{ik\cdot x}[a'b]^{-1} &  u_4&=g^2\int \frac{d^4k}{(2\pi)^4}\Tilde{\Delta}(k)e^{ik\cdot x}[a'b']^{-1}
\end{align*}
Substituting \eqref{i} in \eqref{h} along with the factor given in \eqref{g} we finally obtain
\begin{equation}
\label{j}
    i\mathcal{M}_n=-g^2\int d^4x e^{-iq\cdot x}\Delta(x)\frac{i\chi^{n-1}}{n!}
\end{equation}
Summing over all $n$-loop diagrams, including the tree-level diagram,
\begin{equation}
    \mathcal{M}=\sum_{i=0}^n \mathcal{M}_n=i g^2\int d^4x e^{-iq\cdot x}\Delta(x)\frac{e^{i\chi}-1}{i\chi}
\end{equation}
Notice that the zeroth-order term in $\chi$ above is the tree-level contribution, with the corresponding tree-level diagram
\begin{figure}[h]
    \centering
    \includegraphics[scale=0.8]{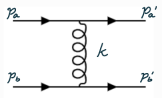}
    \label{fig:my_label}
    \caption{2-2 tree level diagram}
\end{figure}\\

\subsection{Six point scattering amplitude}
We now attempt to calculate the six point scattering amplitude in a similar fashion as done previously for four points. The general $N$ loop diagram for the three particle scattering of $a,b$ \& $c$ will look as follows,
\begin{figure}[h]
    \centering
    \includegraphics[scale=0.75]{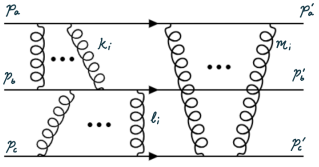}
    \label{fig:my_label}
    \caption{3-3 scattering general ladder diagram}
\end{figure}\\
The contribution of this diagram to the six point scattering amplitude will be
\begin{equation}
\begin{aligned}
    -i\mathcal{M}_{n}=(-ig)^{2n}\sum_{n_k,n_l,n_m}\int \prod_{i=1}^{n_k}\frac{d^4k_i}{(2\pi)^4}\Tilde{\Delta}(k_i)\prod_{j=1}^{n_l}\frac{d^4l_j}{(2\pi)^4}\Tilde{\Delta}(l_j)\prod_{j'=1}^{n_m}\frac{d^4m_{j'}}{(2\pi)^4}\Tilde{\Delta}(m_{j'})\\ \times I\times  (2\pi)^8\delta^4_{s_1}\delta^4_{s_2}
\end{aligned}
\end{equation}
where the number of meson exchanges between $a\&b$, $b\&c$ and $c\&a$ are $n_k$,$n_l$ and $n_m$ respectively, such that $n_k+n_l+n_m=n$. $\delta^4_s$ is the short hand for the momentum conservation relations with product taken over $s_1,s_2\in\{1,2,3\}$ and $s_1\neq s_2$ \footnote{for $n$ particles we will have $(n-1)$ independent momentum conservation relations} for each of the worldlines. The $\delta_s$ are as follows
\begin{equation}
\begin{aligned}
    \delta_1:p_a-p_a'&=q_a=\sum_{i=1}^{n_k} k_i-\sum_{i=1}^{n_m} m_i\\
    \delta_2:p_b-p_b'&=q_b=\sum_{i=1}^{n_l} l_i-\sum_{i=1}^{n_k} k_i\\
    \delta_3:p_c-p_c'&=q_c=\sum_{i=1}^{n_m} m_i-\sum_{i=1}^{n_l} l_i
\end{aligned}
\end{equation}
Now lets choose to eliminate $l_{s_1}$ on the worldline of $b$ using momentum conservation. Hence we will have ($s_1-1$) of $k$s, ($s'_1$) of $l$s to the right of point where $k_{s_1}$ is emmitted. Following the steps in section \ref{sec:worldlines} we obtain the worldline contribution of $b$ to be,
\begin{equation}
\begin{aligned}
     I^{(b) \text{eik}}_{s_1;sym}
    &=\frac{i^{n_k+n_l-1}}{(s_1-1)!(n_l-s_1)!(s'_1)!(n_k-s'_1)!}[b^{(l)}_1b^{(l)}_2\dots b^{(l)}_{s_1-1}]^{-1}[b^{(l)'}_{s_1+1}\dots b^{(l)'}_{n_l}]^{-1}\\
    &\hspace{6cm}\times [b^{(k)}_1b^{(k)}_2\dots b^{(k)}_{s'_1}]^{-1}[b^{(k)'}_{s'_1+1}\dots b^{(k)'}_{n_k}]^{-1} \\
    &=I^{\text{eik}}_{s_1;sym}(l)\times I^{\text{eik}}_{s'_1;sym}(k)
\end{aligned}
\end{equation}
which is essentially equivalent to splitting the family of internal momenta (the '$k$'s in \ref{sec:worldlines}) into two distinct sub families ($k$ and $l$) between which shuffling can no longer be carried out due to loss of indistinguishability. Similarly,
\begin{equation}
\begin{aligned}
\label{l}
     I^{(c) \text{eik}}_{s_2;sym}
    &=\frac{i^{n_l+n_m-1}}{(s_2-1)!(n_l-s_2)!(s'_2-1)!(n_m-s'_2)!}[c^{(l)}_1c^{(l)}_2\dots c^{(l)}_{s_2-1}]^{-1}[c^{(l)'}_{s_2+1}\dots c^{(l)'}_{n_l}]^{-1}\\
    &\hspace{4cm}\times [c^{(m)}_1c^{(m)}_2\dots c^{(m)}_{s'_2}]^{-1}[c^{(m)'}_{s'_2+1}\dots c^{(m)'}_{n_m}]^{-1} \times [c^{(m)}_{s_3}]^{-1}\\
    &=I^{\text{eik}}_{s_2;sym}(l)\times I^{\text{eik}}_{s'_2;sym}(m)\times [c^{(m)}]^{-1}
\end{aligned}
\end{equation}
\begin{equation}
\begin{aligned}
     I^{(a) \text{eik}}_{s_3;sym}
    &=\frac{i^{n_m+n_k-1}}{(s_3-1)!(n_m-s_3)!(s'_3)!(n_k-s'_3)!}[a^{(m)}_1a^{(m)}_2\dots a^{(m)}_{s_3-1}]^{-1}[a^{(m)'}_{s_3+1}\dots a^{(m)'}_{n_m}]^{-1}\\
    &\hspace{6cm}\times [a^{(k)}_1a^{(k)}_2\dots a^{(k)}_{s'_3}]^{-1}[a^{(k)'}_{s'_3+1}\dots a^{(k)'}_{n_k}]^{-1}\\
    &=I^{\text{eik}}_{s_3;sym}(m)\times I^{\text{eik}}_{s'_3;sym}(k)
\end{aligned}
\end{equation}
where we have chosen to eliminate $m_{s_3}$ on the worldline of $a$ via momentum conservation. Notice that in \eqref{l}
we pulled aside the propagator $c^{(m)}_{s_3}$ to which we do not apply the type of eikonal approximation described in \eqref{a}. We also drop the dummy index $s_3$ \footnote{...as this term will be left unpaired because of eliminating its counterpart on the worldline of $a$. We could apply the eikonal approximation, however we will need to account for the symmetry lost making our final expression much less compact.}
\begin{equation}
    c^{(m)}=\frac{1}{(p_c-m_{s_3})^2-M^2-i\epsilon}
\end{equation}
\vspace{2mm}\\
We now glue the three world lines together. We do so by separately applying \eqref{k} the terms corresponding to $k$s, $l$s and $m$s.
\begin{equation}
\begin{aligned}    I^{(k)}&=\sum_{s'_1,s'_3}I^{\text{eik}}_{s'_1;sym}(k)\Pi I^{\text{eik}}_{s'_3;sym}(k)\\
I^{(l)}&=\sum_{s_1,s_2}I^{\text{eik}}_{s_1;sym}(l)\Pi I^{\text{eik}}_{s_2;sym}(l)\\
I^{(m)}&=\sum_{s'_2,s_3}I^{\text{eik}}_{s'_2;sym}(m)\Pi I^{\text{eik}}_{s_3;sym}(m)
\end{aligned}
\end{equation}
Breaking up six point scattering amplitude into the contributions from only $k$,$l$ or $m$,
\begin{equation}
\begin{aligned}
A^{(l)}&=\int\prod_{i=1}^{n_l-1} \frac{d^4l_i}{(2\pi)^4}(-ig)^2\Tilde{\Delta}(l_i)e^{il\cdot x} I^{(l)}\\
&=\frac{(i\chi_l)^{n_l-1}}{(n_l-1)!}
\end{aligned}
\end{equation}
\begin{equation}
\begin{aligned}
A^{(m)}&=\int\prod_{i=1}^{n_m-1} \frac{d^4m_i}{(2\pi)^4}(-ig)^2\Tilde{\Delta}(m_i)e^{im\cdot y} I^{(m)}\\
&=\frac{(i\chi_m)^{n_m-1}}{(n_m-1)!}
\end{aligned}
\end{equation}
\begin{equation}
\begin{aligned}
A^{(k)}&=\int\prod_{i=1}^{n_k} \frac{d^4k_i}{(2\pi)^4}(-ig)^2\Tilde{\Delta}(k_i)e^{ik\cdot (-x-y)} I^{(k)}\\
&=\frac{(i\chi_k)^{n_k}}{(n_k)!}
\end{aligned}
\end{equation}
Where we have followed the notation used in \eqref{i}. Putting everything together along with the $1/n$ factors from \eqref{g} we finally obtain,
\begin{equation}
    i\mathcal{M}^{(1)}_n=-g^2\sum_{n_k,n_l,n_m}\int d^4xd^4ye^{iq_a\cdot x}e^{-iq_b\cdot y}\Delta(x){\Delta}'_c(y)\frac{(i\chi_l)^{n_l-1}}{n_l!}\frac{(i\chi_m)^{n_m-1}}{n_m!}\frac{(i\chi_k)^{n_k}}{(n_k+1)!}
\end{equation}
where we have used
\begin{equation}
    -\frac{c^{(m)}}{m^2-i\epsilon}=\int d^4y \Delta'_c(y)e^{-im\cdot y}
\end{equation}
Here the summation is carried out over $\{n_l-1,n_m-1,n_k\}\in \{0,\dots, (n-1)\}$ such that $n_k+n_m+n_l=n$. Hence the expression is essentially the general term of a multinomial expansion,
\begin{equation}
\begin{aligned}
    \mathcal{M}_n^{(1)}=&-ig^2\int d^4xd^4y e^{iq_a\cdot x}e^{-iq_b\cdot y}\Delta(x){\Delta}'_c(y) \\
    &\times\frac{(\chi_k+\chi_l+\chi_m)^{n-2}-\sum_p(\chi_k+\chi_l)^{n-2}+\sum_p \chi_k^{n-2}}{\chi_k\chi_l\chi_m(n-2)!}
\end{aligned}
\end{equation}
 we have summed over the permutation $p=(klm)$. Finally,
summing over all $n$-loop diagrams, including the tree-level diagram,
\begin{equation}
\mathcal{M}=\sum_{n}\mathcal{M}_n^{(1)}=-ig^2\int d^4xd^4y e^{iq_a\cdot x}e^{-iq_b\cdot y}\Delta(x){\Delta}'_c(y)\frac{e^{i\chi_k}-1}{i\chi_k}\frac{e^{i\chi_l}-1}{i\chi_l}\frac{e^{i\chi_m}-1}{i\chi_m}
\end{equation}
Recall that we have obtained this expression for the case where we have chosen to eliminate an $l$ and an $m$ from the 2 independent conservation relations. However, we could have also alternatively chosen to eliminate $k$ and $m$ or $k$ and $l$. Summing over these contributions we get the final expression for six point scattering amplitude as
\begin{equation}
\label{m}
    \mathcal{M}=-ig^2\int d^4x d^4y \mathcal{T}(x,y)\frac{e^{i\chi_k}-1}{i\chi_k}\frac{e^{i\chi_l}-1}{i\chi_l}\frac{e^{i\chi_m}-1}{i\chi_m}
\end{equation}
where
\begin{equation}
    \mathcal{T}(x,y)=\Delta(x)(e^{iq_a\cdot x}e^{-iq_b\cdot y}{\Delta}'_c(y)+e^{iq_b\cdot x}e^{-iq_c\cdot y}{\Delta}'_a(y)+e^{iq_c\cdot x}e^{-iq_a\cdot y}{\Delta}'_b(y))
\end{equation}
Setting $\chi_k=\chi_l=\chi_m=0$ in \eqref{m} or equivalently looking at the leading order term in the expression, gives us the tree-level amplitude which comes out to be $\mathcal{T}(x,y)$ in the impact parameter space. The tree-level diagrams are as follows
\begin{figure}[h]
    \centering
    \includegraphics[scale=0.75]{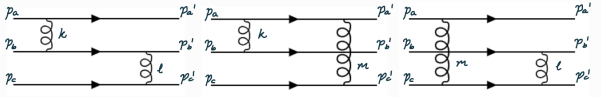}
    \label{fig:my_label}
    \caption{3-3 tree level diagrams}
\end{figure}\\
Hence in the impact parameter space our expression is simply the tree-level amplitude for 3-particle scattering with the $n$-loop corrections as the product of three 2-particle amplitudes.

\subsection{$2N$-point scattering amplitude}
The procedure explained above in order to evaluate the the two- and three- particle amplitudes can be generalised in a straight forward manner to higher points. In the above two cases we have seen that the structure of the two- and three- particle amplitudes in the impact parameter space is directly the product of the tree level diagram with the scattering amplitude contributions of taking two particles at a time out of $n$ particles.
\vspace{2mm}\\
The aforementioned factorization arises due to the symmetric treatment we have used. For any $n$-points the propagator on each of the worldlines can be separated based on the branches between which they mediate interaction due to the  identity \ref{identity}. This lets us treat each such family of momenta independently with the same procedure as that used for two-particle scattering. The only difference arises while gluing the worldlines together, where we treat the terms that arise due to the propagators via which the gluing is performed.
\vspace{2mm}\\
For $n$-particles, $\mathcal{M}$ will be the product of ${}^N C_2$ two-particle scattering amplitudes.The tree-level amplitude will have the following generalised form
\begin{equation}
    \mathcal{M}=-g^{2(N-1)}\int \prod_{i=1}^{N-1}d^4x_i \mathcal{T}(x_1\dots x_{N-1})\prod_{j=1}^{{}^N C_2}\frac{e^{i\chi_j}-1}{\chi_j}
\end{equation}

\section{Conclusions}
Since all $n$-point scattering amplitudes in the eikonal approximation for the case we have studied so far can be written as the product of the tree level diagram with the two-particle scattering
amplitude contributions, the interaction is essentially encoded in the tree level factor. The method outlined above for the case of linearised quantum gravity can be extended to any scalar particle collisions ~\cite{PhysRev.186.1656}.

\section{Future work}
The structure for scattering amplitudes in the eikonal approximation generalised to any n-n scattering process can be used to find the bound states of an n-body problem under the suitable limit. This has been done in ~\cite{Kabat_1992} for the case of 2-2 scattering.


\appendix

\section{The symmetrizing identity}
\label{identity}
Let's consider the function ~\cite{PhysRev.186.1656}
\begin{equation}
    F_n(x)=\sum_{\pi} \frac{1}{a_1+x}\frac{1}{a_1+a_2+x}\dots\frac{1}{a_1+a_2+\dots+a_n+x}
\end{equation}
where the summation is carried out over all permutations of the sequence $(a_1,\dots,a_n)$. Now using the identity
\begin{equation}
    \frac{1}{z}=\int_0^\infty d\alpha e^{-\alpha z} 
\end{equation}
where $Re(z)>0$ we can write $F_n(x)$ as
\begin{equation}
    \frac{1}{x}F_n(x)=\sum_{\pi}\int_0^\infty d\alpha_0\int_0^\infty d\alpha_1\dots\int_0^\infty d\alpha_n e^{-\alpha_0 x}e^{-\alpha_1(a_1+x)}\times \dots\times e^{-\alpha_n(a_1+a_2+\dots+a_n+x)}
\end{equation}
Performing the variable change $\beta_0=\alpha_0+\alpha_1+\dots+\alpha_n$, $\beta_1=\alpha_1+\alpha_2+\dots+\alpha_n$ and so on till $\beta_n=\alpha_n$.
\begin{equation}
\begin{aligned}
\label{A}
    \frac{1}{x}F_n(x)=\sum_{\pi}{\iint\dots\int}_{\beta_0\geq\beta_1\geq\dots\geq\beta_n\geq 0}d\beta_0 d\beta_1\dots d\beta_n \\
    \times e^{-\beta_0 x}e^{-\beta_1 a_1}\dots e^{-\beta_n a_n}
\end{aligned}    
\end{equation}
However as we are summing over all permutations, for any one of these permutation we have the general relation,
\begin{gather}
    \beta_i\rightarrow\beta_{\pi(i)}\\
    \beta_0\geq\beta_{\pi(1)}\geq\dots\geq \beta_{\pi(n)}\geq 0
\end{gather}
and so summing over all permutations implies that for any $\beta_{\pi(i)}$ 
\begin{equation}
    \beta_0\geq\beta_{\pi(i)}\geq0
\end{equation}
Hence \eqref{A} reduces to an integral over a hypercube with side length $\beta_0$.
\begin{equation}
\begin{aligned}
    \frac{1}{x}F_n(x)&=\int_0^{\infty}d\beta_0 e^{-\beta_0 x}\int_0^\beta d\beta_1 \dots\int_0^\beta d\beta_n e^{-\beta_1 a_1}\dots e^{-\beta_n a_n}\\
    \implies F_{n}(x)&=x\int_0^\infty d\beta_0 e^{-\beta_0 x}\prod_{i=1}^n\left(\frac{1-e^{-\beta a_i}}{a_i}\right)
\end{aligned}
\end{equation}
Applying integration by parts,
\begin{equation}
     F_{n}(x)=\int_0^\infty d\beta_0 e^{-\beta_0 x}\frac{\partial}{\partial\beta_0}\prod_{i=1}^n\left(\frac{1-e^{-\beta a_i}}{a_i}\right)
\end{equation}
Therefore for $x=0$,
\begin{equation}
    F_{n}(0)=\prod_{i=1}^n\left(\frac{1-e^{-\beta a_i}}{a_i}\right)\bigg\rvert_{\beta_0=0}^{\beta_0=\infty}=\frac{1}{a_1a_2\dots a_n}
\end{equation}


 \bibliographystyle{JHEP}
 \bibliography{biblio.bib}



\end{document}